\documentclass{article}
\usepackage{arxiv}

\usepackage[utf8]{inputenc}
\usepackage[T1]{fontenc}
\usepackage{cite}
\usepackage{amsmath,amssymb,amsfonts}
\usepackage{graphicx}
\usepackage{textcomp}
\usepackage{xcolor}
\usepackage{booktabs}
\usepackage{multirow}
\usepackage{array}
\usepackage{url}
\usepackage{hyperref}
\usepackage{float}
\usepackage{placeins}
\usepackage{microtype}
\usepackage{algorithm}
\usepackage{algpseudocode}

\graphicspath{{figures/}{./}}

\hypersetup{
  colorlinks=true, linkcolor=blue, citecolor=blue, urlcolor=blue,
  pdfauthor={Eric Yocam, Varghese Vaidyan, Micah Flack, Gurcan Comert, Judith Mwakalonge},
  pdftitle={Side-Channel Attacks Survive Noise Cancellation in 3D Printers},
  pdfsubject={Computer Science - Cryptography and Security},
  pdfkeywords={side-channel attacks, additive manufacturing, vibration analysis, noise cancellation, print duration, cyber-physical security}
}

\newcommand{\codeurl}[1]{\url{#1}}

\begin{document}

\title{Side-Channel Attacks Survive Noise Cancellation in 3D Printers}

\author{
  Eric Yocam \\ Department of Computer Science\\ California Polytechnic State University\\ San Luis Obispo, CA 93407, USA
  \And
  Varghese Vaidyan \\ Beacom College of Computer and Cyber Sciences\\ Dakota State University\\ Madison, SD 57042 USA
  \And
  Micah Flack \\ Idaho National Laboratory
  \And
  Gurcan Comert \\ Department of Computational Data Science and Engineering\\ North Carolina A\&T State University\\ Greensboro, NC 27411, USA
  \And
  Judith L. Mwakalonge \\ Department of Engineering\\ South Carolina State University\\ Orangeburg, SC 29117, USA
}

\date{July 2026}
\maketitle

\begin{abstract}
Active Motor Noise Cancellation (AMNC) is a noise-reduction feature shipped in
commercial fused deposition modeling (FDM) 3D printers. Because it suppresses the
acoustic emissions that side-channel attacks exploit, it has security-relevant side
effects, though we find no evidence it was designed as a security control. We present
a duration-controlled evaluation of side-channel leakage on AMNC-equipped hardware,
using a public dataset of 144 synchronized acoustic and vibration recordings from two
Bambu Lab printers across 12 object classes. Spectral analysis confirms that
suppression is measurably active: the motor-resonance band rises only 4.92\,dB above
its background-relative baseline during printing. Leakage nonetheless survives it.
Acoustic classification is at chance for 30-second observation windows (11.11\%,
permutation \(p=0.188\)) but reaches 27.08\% on a duration-clean 60-second window and
40.28\% when sampling is distributed across the print, against an 8.33\%
baseline---the channel is suppressed within short windows, not eliminated. The
dominant discriminator, however, is print duration itself, which classifies at 63.89\%
(95\% CI [55.78, 71.28]) from recording length alone and is validated against sliced
G-code print time at \(r=0.907\). Vibration carries genuine geometry information
independent of duration: with the observation window equalized by truncation,
classification reaches 45.83\% against a 25\% baseline (permutation \(p=0.023\)). It
nonetheless adds no measurable information beyond duration in paired comparison. The
leak is architecture-specific, present on the core-XY device (36.11\%) and
indistinguishable from chance on the bed-slinger (13.89\%, CI lower bound 7.72\%), and
a consumer handset recovers as much as a mounted accelerometer (25.00\% versus
26.39\%). We conclude that noise cancellation raises the observation time an acoustic
attacker requires without eliminating the channel, and leaves print duration---the
strongest discriminator---entirely untouched. We release all code and analysis
artifacts.
\end{abstract}


\section{Introduction}

Additive manufacturing has grown into a multi-billion-dollar industry, and FDM
technology now underpins distributed manufacturing ecosystems in which design
files cross organizational boundaries to remote or third-party
printers~\cite{yampolskiy2016,yampolskiy2018}. Even when files are encrypted
in transit and at rest, the physical printing process leaks information through
unintended emanations observable without cyber
access~\cite{faruque2016,chhetri2019,zeltmann2016}.

Side-channel attacks (SCAs) on 3D printers exploit acoustic noise, mechanical
vibration, electromagnetic radiation, power draw, and thermal emission to
identify or reconstruct printed
geometry~\cite{backes2010,al_faruque2016_smartphone,gupta2020}. Al~Faruque et
al.~\cite{faruque2016} demonstrated 78.35\% acoustic axis prediction accuracy;
Gatlin et al.~\cite{gatlin2021} achieved over 99\% via the power side channel;
Jamarani et al.~\cite{jamarani2025} reached 98.80\% by fusing acoustic and
magnetic channels; Garza et al.~\cite{garza2025} reconstructed G-code from
optical video.

Bambu Lab deploys Active Motor Noise Cancellation (AMNC) in their A1~Mini and P1P
printers, suppressing motor resonance frequencies via adaptive current
control~\cite{bambulab2023,notebookcheck2023}. We treat AMNC throughout as a
noise-reduction feature: we located no patent, firmware documentation, or engineering
statement connecting it to emanation suppression, and we make no claim regarding
vendor design intent. Its relevance here is functional rather than intentional.
Because AMNC attenuates precisely the acoustic emissions that prior side-channel work
exploits~\cite{faruque2016,chhetri2017_ccs}, it is the first opportunity to ask what a
deployed acoustic suppression mechanism does to side-channel leakage in practice. Its
scope is airborne acoustic output; it does not act on structural vibration, power, or
electromagnetic emanation~\cite{quietprint2026}.

This paper asks two questions: \textit{does an active acoustic suppression mechanism
eliminate the channel it attenuates, and what dominates leakage once confounds are
controlled?} We answer empirically using the first public dataset collected on
AMNC-equipped printers~\cite{madamopoulos2024,madamopoulos2024_zenodo}. A central
methodological finding shapes the answer: print duration in this dataset is so
strongly associated with object class that it discriminates better than any sensor
channel, and results reported without controlling for it may partly reflect elapsed
time rather than geometry. Our contributions are:

\begin{enumerate}
\item \textbf{Acoustic leakage survives active suppression.} Spectral analysis
confirms suppression is operating (motor band elevated only 4.92\,dB above its
background-relative baseline). Classification is nonetheless at chance only for short
observation windows: 11.11\% at 30\,s (permutation \(p=0.188\)), rising to 27.08\% at
a duration-clean 60\,s window and 40.28\% under distributed sampling. Suppression
raises the observation time an attacker needs; it does not remove the channel.

\item \textbf{Print duration is the dominant discriminator.} Recording length alone
classifies the object at 63.89\% (95\% CI [55.78, 71.28]) against an 8.33\% baseline,
and tracks sliced G-code print time at Pearson \(r=0.9073\) (\(n=144\)). No acoustic
countermeasure addresses this channel, and an adversary obtains it from start and stop
times alone.

\item \textbf{Duration-controlled characterization of the vibration channel.} With the
observation window equalized by truncation, vibration classifies at 45.83\% against a
25\% baseline (permutation \(p=0.023\)), demonstrating geometry-dependent leakage
independent of elapsed time. In paired comparison, however, vibration adds no
measurable information beyond duration.

\item \textbf{Equivalence of capture configurations.} A consumer handset application
(25.00\%) performs comparably to a mounted accelerometer (26.39\%), lowering the
physical access an adversary requires.

\item \textbf{Device specificity.} Vibration leakage is present on the core-XY P1P
(36.11\%) and not distinguishable from chance on the bed-slinger A1~Mini (13.89\%,
95\% CI [7.72, 23.71]); cross-device transfer is at chance in both directions.

\item \textbf{A bounded security conclusion.} The demonstrated capability is closed-set
identification, not reconstruction of unknown geometry. Acoustic suppression is an
incomplete control against the channel it targets and no control at all against print
duration.
\end{enumerate}

We compare this work against prior 3D printer SCA studies in
Section~\ref{sec:related} (Table~\ref{tab:priorwork}).

\section{Related Work}
\label{sec:related}
A growing body of work shows that 3D printers leak through multiple physical
channels. We review them by channel and by mitigation, then summarize the
landscape in Table~\ref{tab:priorwork}.

\subsection{Acoustic, Power, and Electromagnetic Channels}
The earliest acoustic SCA on a printing system targeted dot-matrix
printers~\cite{backes2010}. Al~Faruque et al.~\cite{faruque2016} extended
acoustic SCAs to FDM printers; Chhetri et al.~\cite{chhetri2017_ccs} showed
that smartphone-based attacks are feasible. Kubiak et al.~\cite{kubiak2020} and
Sta{\'n}czak et al.~\cite{stanczak2021} studied vibration--acoustic shape
reconstruction, establishing vibration as an independent leakage channel.
Gatlin et al.~\cite{gatlin2021} reconstructed geometry via stepper-motor
current; Dolgavin et al.~\cite{dolgavin2025} extended power SCAs to industrial
PBF machines; Belikovetsky et al.~\cite{chhetri2018_tifs} proposed audio
signatures for print integrity. Kocher et al.~\cite{kocher2011} and Genkin et
al.~\cite{genkin2014} established the broader feasibility of power and acoustic
physical-channel attacks.

\subsection{Multi-Modal, Optical, and Thermal Channels}
Costa et al.~\cite{costa2021} released a multi-channel dataset from an
Ultimaker~3, with the vibration channel independently achieving strong state
estimation. Jamarani et al.~\cite{jamarani2025} fused acoustic and magnetic
channels; Streit et al.~\cite{streit2023} demonstrated electromagnetic SCA of
neural accelerators; Garza et al.~\cite{garza2025} reconstructed G-code from
optical video; Chhetri et al.~\cite{chhetri2016_thermal} pioneered thermal
forensics; Liang and Beyah~\cite{liang2023} proposed optical defenses.

\subsection{Mitigations, Integrity, and Evaluation}
QuietPrint~\cite{quietprint2026} proposed a G-code modification as an acoustic
defense; Chhetri et al.~\cite{chhetri2017_fix} proposed leakage-aware G-code
generation; Al~Faruque et al.~\cite{alfaruque_patent2019} patented a physical
process encryption. Others studied Trojans~\cite{pearce2022}, blockchain
protection~\cite{shi2021}, physical hashing~\cite{brandman2023}, MitM
attacks~\cite{alkofahi2024}, firmware attacks~\cite{rais2024}, ML-based
detection~\cite{wu2022}, authentication~\cite{mouris2024}, and broad
surveys~\cite{tsoutsos2020,zeltmann2016,belikovetsky2017,gupta2020,xing2021}.

Two gaps motivate this study. First, no prior work characterizes side-channel leakage
on hardware where an acoustic suppression mechanism is active during collection.
Second, and more consequentially for the field, we are not aware of prior 3D printer
SCA work that controls for print duration. Object sets in this literature routinely
span an order of magnitude in print time, and duration is trivially observable;
results reported against a chance baseline rather than a duration baseline may
therefore overstate geometric leakage. Table~\ref{tab:priorwork} records this column
explicitly.

\begin{table}[H]
\caption{Prior 3D printer SCA research: methodology and scope.}
\label{tab:priorwork}
\centering
\renewcommand{\arraystretch}{1.05}
\begin{tabular}{lllll}
\toprule
\textbf{Study} & \textbf{Channel(s)} & \textbf{Printer} &
\textbf{Duration ctrl.} & \textbf{Dataset} \\
\midrule
Al Faruque~\cite{faruque2016} (2016) & Acoustic & FDM & No & Private \\
Chhetri~\cite{chhetri2017_ccs} (2016) & Acoustic & FDM & No & Private \\
Kubiak~\cite{kubiak2020} (2020) & Acoustic & FDM & No & Private \\
Gatlin~\cite{gatlin2021} (2021) & Power & FDM & No & Private \\
Costa~\cite{costa2021} (2021) & Multi & Ultimaker 3 & No & Public \\
Jamarani~\cite{jamarani2025} (2025) & Acous.+Mag. & FDM & No & Private \\
Dolgavin~\cite{dolgavin2025} (2025) & Power & PBF & No & Private \\
Garza~\cite{garza2025} (2025) & Optical & FDM & No & Private \\
QuietPrint~\cite{quietprint2026} (2026) & Acoustic & FDM & No & Private \\
\midrule
\textbf{This work} (2026) & \textbf{Acous.+Vib.} & \textbf{Bambu P1P, A1 Mini}
  & \textbf{Yes} & \textbf{Public} \\
\bottomrule
\end{tabular}
\end{table}
\FloatBarrier

\section{Threat Model and Experimental Design}

\subsection{Adversary}
The adversary seeks to \emph{identify} which of a set of known designs is being
printed (closed-set classification), operating in physical proximity with one
or more passive sensors and no cyber access~\cite{faruque2016,chhetri2017_ccs,
jamarani2025}. A labeled corpus from known print runs on the target device
enables supervised classification~\cite{gatlin2021,quietprint2026}. We
distinguish identification from \emph{reconstruction} (recovering unknown geometry
or G-code) and claim only the former.

\subsection{Capture Configurations and Adversary Capability}
\label{sec:capture}
The dataset comprises two independent capture configurations, balanced six and six
across every object class. Seventy-two recordings were captured with a Teensy~4.0
microcontroller and an attached accelerometer, at a measured vibration sampling rate
of 500.0\,Hz. Seventy-two were captured with an iPhone application, at a measured rate
of 199.6\,Hz. Both record audio at 44.1\,kHz.

The two configurations imply materially different adversaries. The Teensy
configuration requires a sensor physically mounted to the target machine: it presumes
access sufficient to install hardware, and the installed sensor is discoverable on
inspection. The iPhone configuration requires only a consumer handset in proximity,
installs nothing on the printer, and is not distinguishable from ordinary presence.
Section~\ref{sec:persystem} shows the two recover comparable information, so the
weaker access assumption is the operative one.

The public release does not document accelerometer mounting position, microphone
standoff distance, or axis orientation. We state this rather than infer it, and scope
our claims to what the data supports. Standoff attenuation and through-surface
coupling are untested here.

\subsection{Constraint on Causal Inference}
AMNC is active in hardware throughout data collection and neither device exposes a
user-accessible control to disable it. No causal statement about AMNC is therefore
supportable from this dataset. We report what leakage is present while suppression is
operating, and we use the dataset's own background noise recordings
(Section~\ref{sec:spectral}) to establish that suppression is measurably active. We do
not claim to measure what suppression \emph{reduces}, only that leakage persists in
its presence. We state this before presenting any result so that no associative
finding is mistaken for a causal one.

\subsection{Analyses}
We evaluate: (1) the acoustic channel as a function of observation window; (2) print
duration as a discriminator, validated against G-code ground truth; (3) vibration with
summary features, under equalized observation windows; (4) a feature ablation
separating amplitude from frequency content; (5) duration controls (T1--T3); (6) a
full-sequence temporal model with an order-shuffle control; (7)
per-capture-configuration and per-printer analyses, and cross-printer transfer; and
(8) spectral verification of suppression.

\section{Dataset and Feature Extraction}

\subsection{Dataset}
We use the Madamopoulos--Tsoutsos 2024
dataset~\cite{madamopoulos2024,madamopoulos2024_zenodo} (Zenodo DOI:
10.5281/zenodo.13329934, CC~BY~4.0): synchronized audio and triaxial accelerometer
captures on a P1P (core-XY) and an A1~Mini (bed-slinger), covering 12 objects with six
paired recordings per object per printer (144 audio--vibration pairs total). Audio
durations span 88.3\,s to 11{,}273.3\,s and the paired vibration captures 116.9\,s to
11{,}265.5\,s, a factor of roughly 100 in each case; both are strongly associated with
object class---the central methodological fact of this study. Where a duration-clean
window is claimed below it is assessed against the shorter of the two, so no recording
is truncated. The release also includes the sliced G-code for every object, which we
use as ground truth for print time (Section~\ref{sec:duration}).

\subsection{Recording Pairing and Cross-Validation}
\label{sec:pairing}
Each recording session stores its audio alongside its own accelerometer capture. We
pair each audio file with the capture in its own session folder, giving 144 distinct
audio--vibration pairs, and treat each recording as a single labeled sample. An audit
of the catalog establishes that the 144 recordings occupy 144 distinct sessions, one
sample per session; grouped cross-validation is therefore identical to stratified
cross-validation on this dataset, and we describe the protocol accurately as
stratified five-fold cross-validation. Catalog composition (144 recordings, 12
classes, 2 printers) is verified programmatically before any analysis executes.
Algorithm~\ref{alg:eval} states the evaluation protocol.

\begin{algorithm}[H]
\caption{Duration-controlled evaluation protocol.}
\label{alg:eval}
\begin{algorithmic}[1]
\Require recordings $\mathcal{R}$; each $r$ has audio $a_r$, vibration capture $c_r$, label $y_r$
\State \textbf{Verify} catalog composition; abort on mismatch
\State \textbf{Pair} each $a_r$ with the capture $c_r$ in its session folder
\State \textbf{Measure} sampling rate $f_r$ from the timestamp channel
\State \textbf{Truncate} each $c_r$ to a common observation window in seconds
\State \textbf{Extract} feature vector $x_r \gets \textsc{Features}(c_r, f_r)$
\State \textbf{Validate} with stratified $k$-fold cross-validation
\State \textbf{Assess} against a duration baseline, Wilson CIs, a label-permutation
       null, paired McNemar tests, and---for sequence models---an order-shuffle control
\State \Return accuracy and validity statistics
\end{algorithmic}
\end{algorithm}
\FloatBarrier

\subsection{Feature Extraction}
\textbf{Acoustic.} Audio is loaded at 16~kHz; we extract 13
MFCCs~\cite{davis1980}, spectral centroid, bandwidth, rolloff, and
zero-crossing rate (mean and std), giving a 32-d vector. The observation window is a
controlled variable (Section~\ref{sec:acoustic}) rather than a fixed parameter. We
additionally apply a simulated notch bank at 120--360~Hz
($Q\!\in\!\{15,30,60\}$) on top of the active hardware mechanism.

\textbf{Vibration (summary).} Per axis we compute mean, std, RMS, peak-to-peak, and
dominant FFT frequency, giving a 15-d vector per recording. Dominant frequency is
computed in hertz using each recording's measured sampling rate; because the dataset
mixes two capture configurations at different rates, normalized-frequency features are
not comparable across recordings.

\textbf{Vibration (temporal).} Each recording is divided into $T\!=\!120$ ordered
segments of equal duration in seconds, identical across recordings, giving an ordered
$120\times12$ sequence, z-scored per recording to remove absolute amplitude and
isolate temporal shape.

\section{Classification and Evaluation}
Summary-feature attacks use a Random Forest (200 trees) with standardized features and
stratified five-fold cross-validation. The temporal attack uses a dilated 1D-CNN over
the full 120-step sequence, averaged over five seeds with deterministic algorithm
selection.

Significance uses Wilson confidence intervals~\cite{wilson1927} and a
label-permutation null. For the permutation test, class labels are permuted and the
full cross-validation pipeline is re-executed inside each permutation so that no fitted
model leaks across the null; $B=1000$ permutations are used; the estimator is
$(r{+}1)/(B{+}1)$, bounded below by $0.001$; and the test is one-sided against the
observed accuracy. Where two feature sets are compared on identical folds we use an
exact McNemar test on the paired predictions, which is the appropriate test for
whether one feature set adds information to another. All analyses use deterministic
data ordering and fixed random seeds.

\section{Results}
\label{sec:results}

\subsection{Acoustic Channel: Observation Window}
\label{sec:acoustic}
\begin{figure}[H]
\centering
\includegraphics[width=0.8\textwidth]{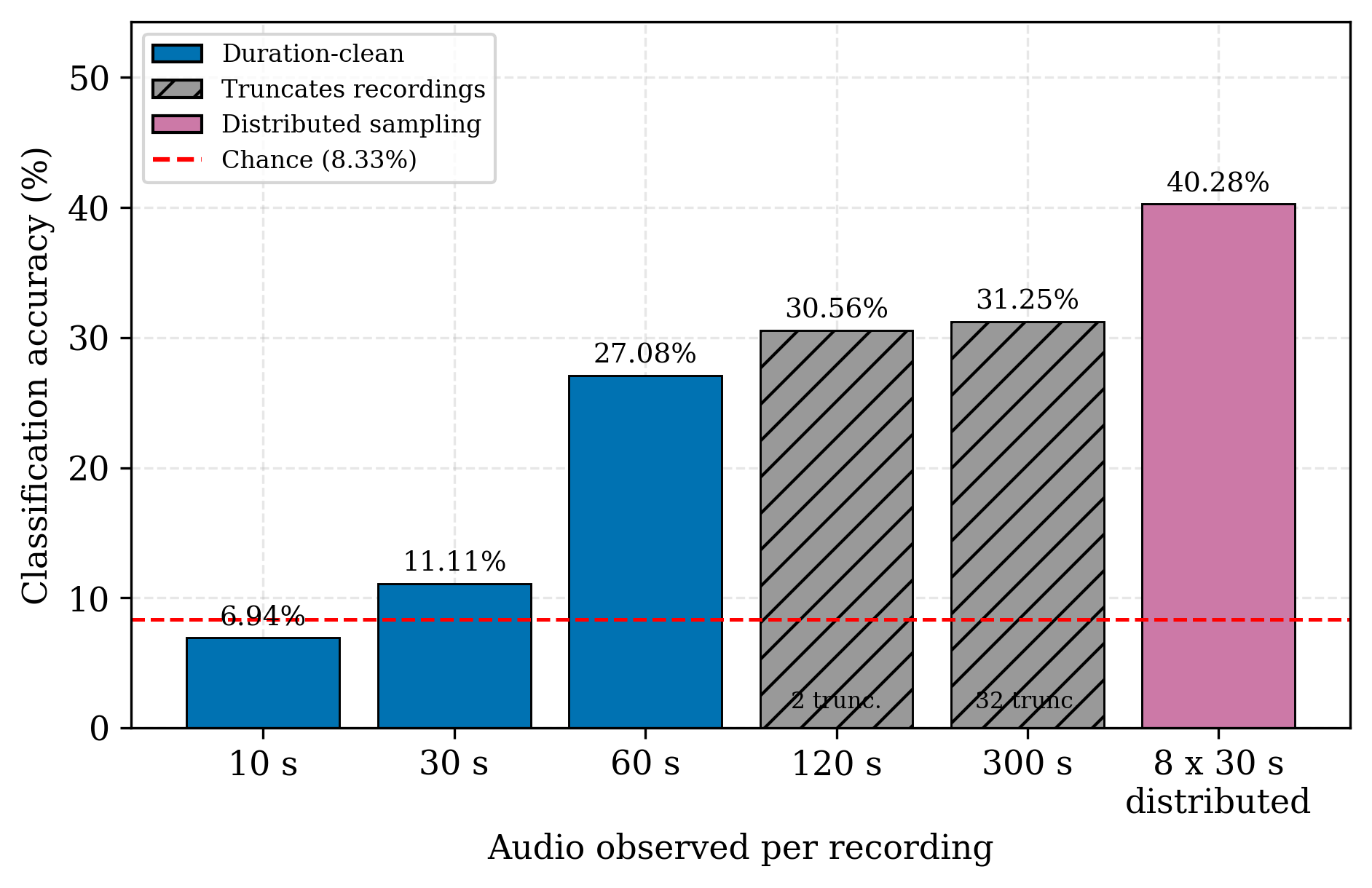}
\caption{Acoustic accuracy versus observation window. The 60-second window is
duration-clean; longer windows truncate short recordings and are shown hatched.}
\label{fig:window}
\end{figure}

Acoustic classification depends strongly on how much audio the adversary observes
(Table~\ref{tab:window}, Figure~\ref{fig:window}). At the 30-second window, accuracy
is 11.11\% (Wilson 95\% CI [6.96, 17.29], permutation
\(p=0.188\))---indistinguishable from the 8.33\% baseline. At 60 seconds it reaches
27.08\%. The 60-second window is duration-clean: the shortest recording is 88.3\,s, so
every sample supplies a full, equal-length window and no truncation artifact is
possible. Distributing eight 30-second windows across the print raises accuracy to
40.28\%.

The 120\,s and 300\,s windows truncate 20 and 36 recordings respectively; they are
reported for completeness and are not used to support claims. The simulated notch bank
leaves accuracy near baseline across quality factors (13.89\%, 13.19\%, 13.19\% for
$Q=15,30,60$).

The null result at 30 seconds is therefore genuine for a 30-second attack, but it is
not a property of the channel. Acoustic suppression raises the observation time an
attacker requires; it does not eliminate the channel.

\begin{table}[H]
\caption{Acoustic accuracy by observation window, with truncation audit.}
\label{tab:window}
\centering\renewcommand{\arraystretch}{1.10}
\begin{tabular}{lrr}
\toprule
\textbf{Audio observed} & \textbf{Accuracy (\%)} & \textbf{Recordings truncated} \\
\midrule
10\,s                     & \phantom{0}6.94 & 0 \\
30\,s                     & 11.11 & 0 \\
\textbf{60\,s}            & \textbf{27.08} & \textbf{0} \\
120\,s                    & 30.56 & 20 \\
300\,s                    & 31.25 & 36 \\
$8\times30$\,s distributed & 40.28 & --- \\
\bottomrule
\end{tabular}
\end{table}
\FloatBarrier

\FloatBarrier

\subsection{Print Duration as a Side Channel}
\label{sec:duration}
Recording length alone classifies the printed object at 63.89\% (Wilson 95\% CI
[55.78, 71.28]) against an 8.33\% baseline---higher than any sensor channel examined
in this study.

\begin{figure}[H]
\centering
\includegraphics[width=0.62\textwidth]{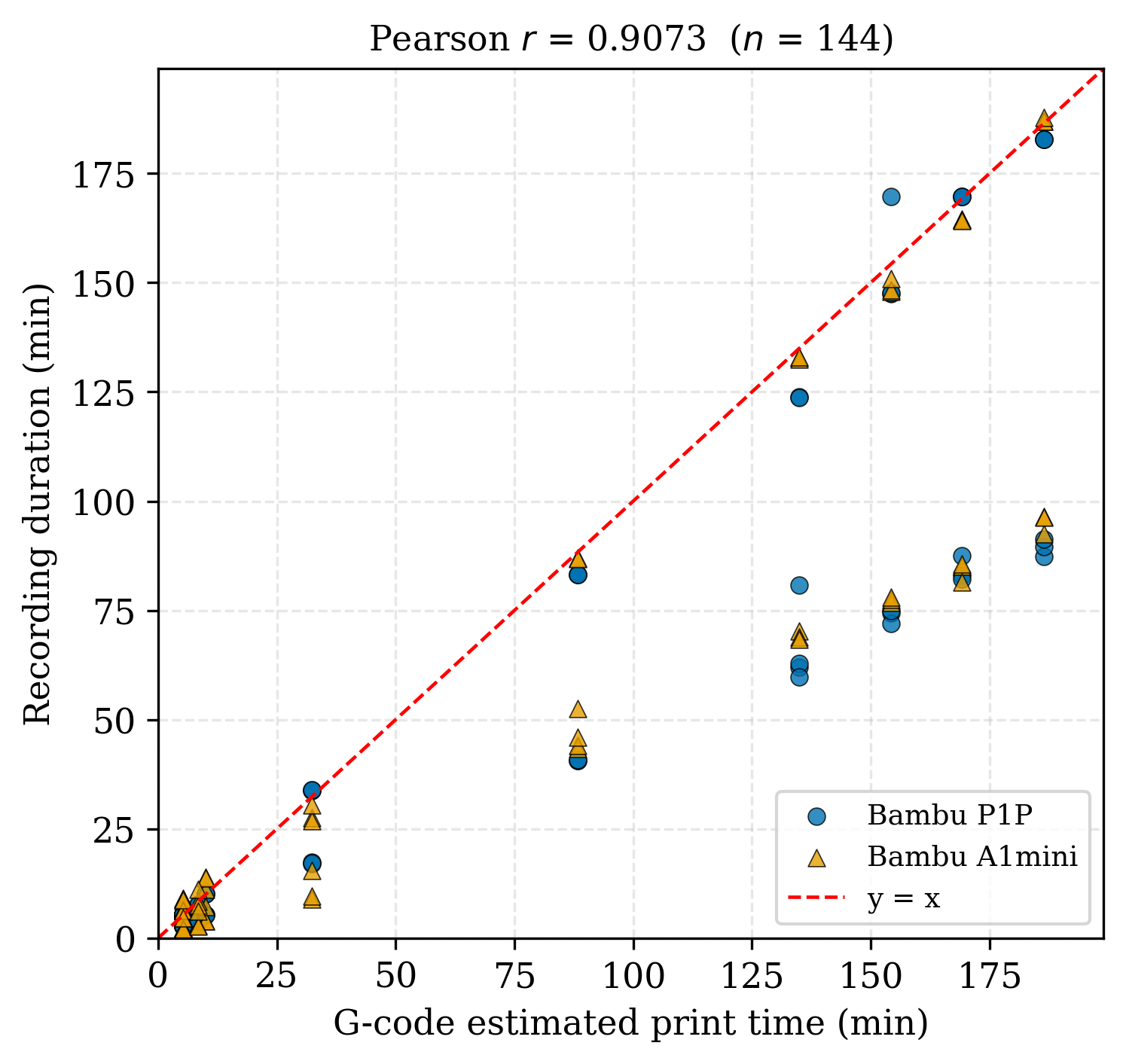}
\caption{Recording duration against sliced G-code estimated print time
(\(r=0.9073\), \(n=144\)). Dashed line is \(y=x\).}
\label{fig:gcode}
\end{figure}

To establish that this reflects genuine print duration rather than a recording
artifact, we parsed the estimated print time from the sliced G-code distributed with
the dataset. Recording duration correlates with G-code print time at Pearson
\(r=0.9073\) (Spearman \(\rho=0.8933\), \(n=144\)); print times encoded in the G-code
filenames give \(r=0.9075\) (Figure~\ref{fig:gcode}). Recording length tracks the
physical process.

This channel requires no sensor on or near the machine. An adversary who observes only
when a print begins and ends obtains it.

\FloatBarrier

\subsection{Vibration Channel under Equalized Observation}
\begin{figure}[H]
\centering
\includegraphics[width=0.75\textwidth]{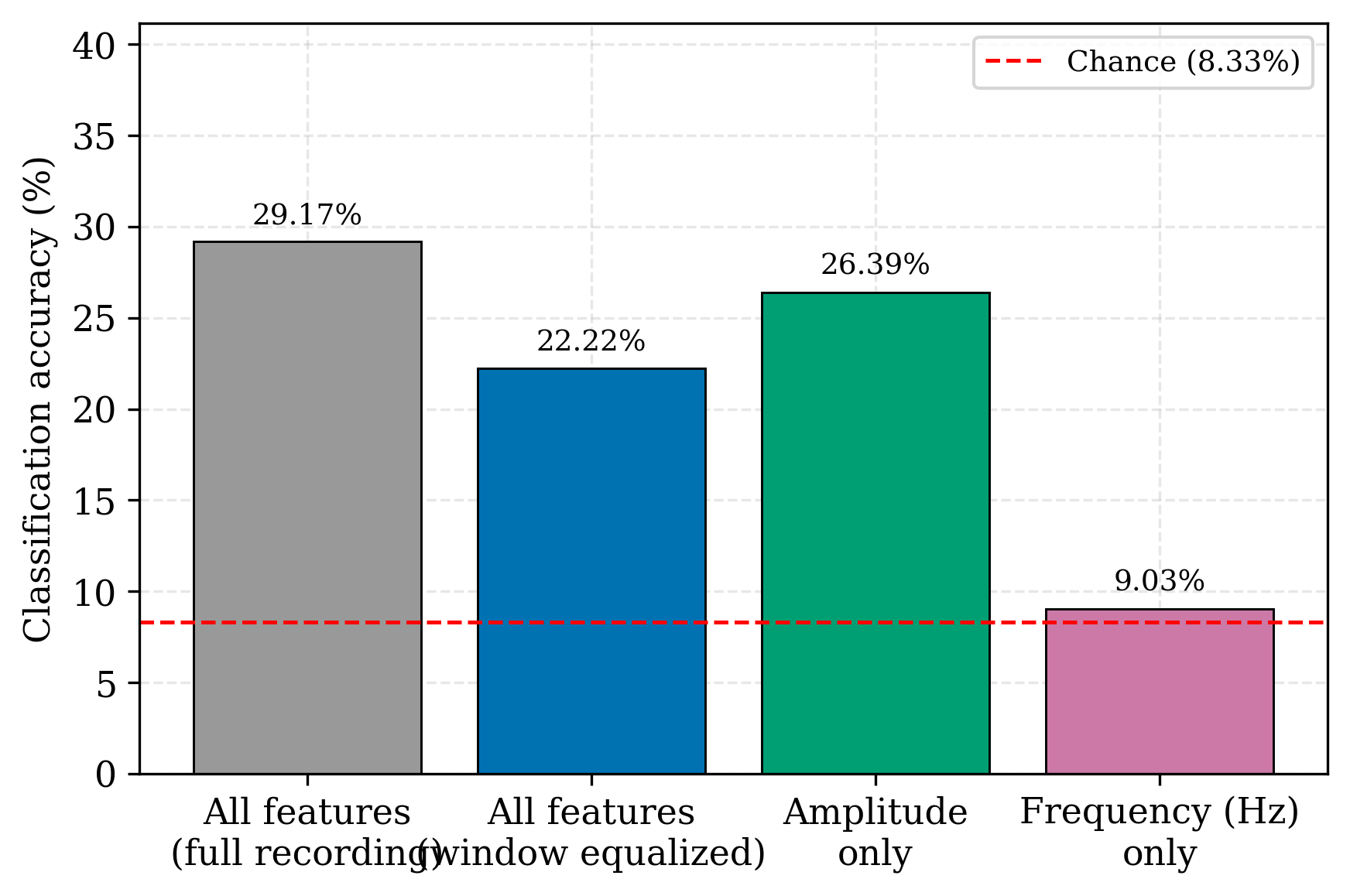}
\caption{Feature ablation. Amplitude features carry the signal; frequency features
computed in hertz are at chance within the observable bandwidth.}
\label{fig:ablation}
\end{figure}

Pooled across both printers and both capture configurations, vibration summary
features classify at 29.17\% (CI [22.36, 37.05]) on full recordings. Equalizing the
observation window across all recordings reduces this to 22.22\% (CI [16.20, 29.68]),
still well above the 8.33\% baseline. A feature ablation localizes the signal to
amplitude (26.39\%) rather than waveform shape: frequency-only features, computed in
hertz, sit at 9.03\% (CI [5.35, 14.83]), straddling baseline
(Figure~\ref{fig:ablation}).

\FloatBarrier

\subsection{Duration Controls}
\label{sec:controls}
Because duration is the strongest discriminator, every other channel must be assessed
against it rather than against chance. We report three controls
(Table~\ref{tab:controls}).

\textbf{T1: incremental contribution.} Within the long-duration stratum, where
duration alone classifies at 95.83\% (CI [86.02, 98.85]), adding sensor features never
improves on duration-only predictions. Across every combination tested, the number of
samples that duration classified incorrectly and the combined feature set classified
correctly is zero; combined accuracies fall to 79.17\% (vibration), 56.25\%
(acoustic), and 50.00\% (both). Vibration and acoustic features add no measurable
information beyond duration. We attribute the decrease to feature dilution rather than
to destruction of information.

\textbf{Fixed-offset acoustic window.} Because window position covaries with total
print length, we also evaluated acoustic features from a fixed 60--120\,s offset
window, identical for every recording and therefore independent of print duration.
Within the long-duration stratum this yields 31.25\% (CI [19.95, 45.33]) against a
25\% baseline. The interval includes chance, so once window position is held fixed the
acoustic channel does not clearly discriminate within the stratum.

\textbf{T2: equalized observation window.} Truncating long-stratum recordings to a
uniform 543\,s, which equalizes the observation window by construction, yields
vibration accuracy of 45.83\% (CI [32.58, 59.71], permutation \(p=0.023\)) against a
25\% baseline. This is our cleanest evidence that vibration carries
geometry-dependent information independent of elapsed time. Notably, accuracy does not
degrade under truncation, which excludes a minimum-runtime explanation for the
leakage.

\textbf{T3: duration-overlap subset.} We also constructed a duration-matched subset by
quantile binning. This control was inconclusive: duration alone still classified the
resulting subset at 60.34\%, so binning did not eliminate duration separation. We
report it for completeness and do not rely on it.

Taken together: vibration carries genuine geometry information (T2), and duration is
nonetheless the stronger readout of the same underlying process (T1). Both statements
are supported, and they are not in conflict---geometry determines both what the
machine does and how long it takes.

\begin{table}[H]
\caption{Duration controls. Chance is 25\% within strata.}
\label{tab:controls}
\centering\renewcommand{\arraystretch}{1.10}
\begin{tabular}{llr}
\toprule
\textbf{Control} & \textbf{Condition} & \textbf{Acc.\ (\%)} \\
\midrule
T1 & Duration only                  & 95.83 \\
T1 & Duration $+$ vibration         & 79.17 \\
T1 & Duration $+$ acoustic          & 56.25 \\
T1 & Duration $+$ both              & 50.00 \\
\midrule
T2 & Vibration, window equalized    & \textbf{45.83} \\
\midrule
T3 & Duration only (control failed) & 60.34 \\
\bottomrule
\end{tabular}
\end{table}
\FloatBarrier

\subsection{Capture Configuration}
\label{sec:persystem}
The two capture configurations recover comparable information: the iPhone application
reaches 25.00\% (CI [16.44, 36.09]) and the mounted Teensy accelerometer 26.39\% (CI
[17.59, 37.58]). The intervals overlap substantially. For the iPhone configuration,
combining duration and vibration yields exactly the duration-only accuracy (59.72\%),
consistent with T1.

Instrumented hardware confers no advantage over a consumer handset in proximity. The
practical access requirement for this attack is correspondingly lower than the
mounted-sensor framing implies.

\subsection{Device Architecture and Transfer}
\begin{figure}[H]
\centering
\includegraphics[width=0.8\textwidth]{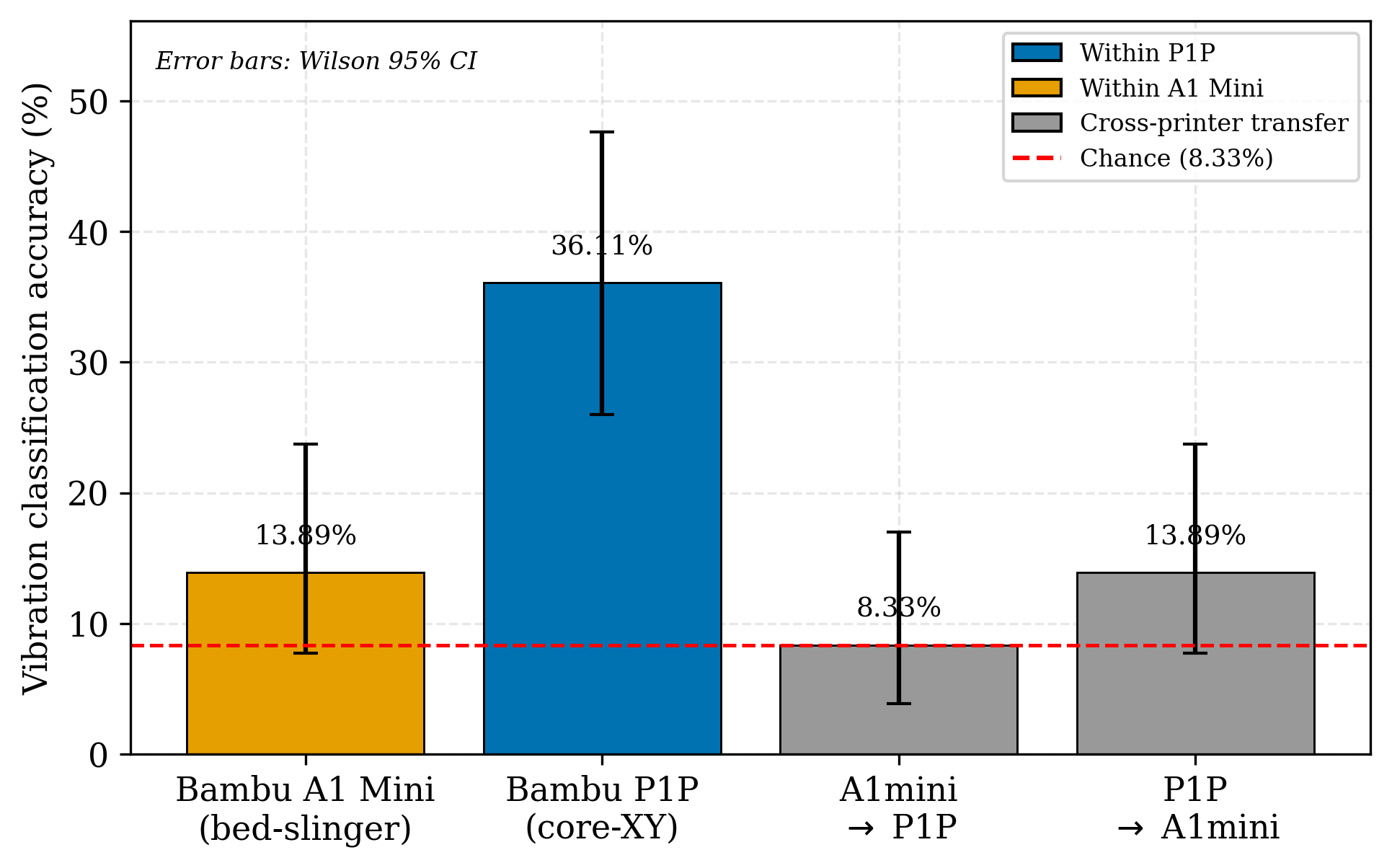}
\caption{Per-printer vibration accuracy and cross-printer transfer.}
\label{fig:cross}
\end{figure}

Vibration leakage is architecture-specific. On the core-XY P1P it reaches 36.11\% (CI
[25.98, 47.65]); on the bed-slinger A1~Mini it reaches 13.89\% with a confidence
interval lower bound of 7.72\%, below the 8.33\% baseline---on that machine vibration
is not distinguishable from chance in this dataset. Cross-printer transfer is at
chance in both directions (8.33\% and 13.89\%), confirming that vibration signatures
do not generalize across architectures (Figure~\ref{fig:cross}).

\FloatBarrier

\subsection{Temporal Structure}
A full-sequence model over equal-duration segments reaches 25.42\%~$\pm$~2.31 against
15.28\%~$\pm$~4.33 when segment order is destroyed (five seeds). The 10.14-point gap
indicates a genuine sequential component tied to print progression. We note that the
shuffled condition remains above the 8.33\% baseline, so some non-temporal structure
survives per-recording normalization, and that with five seeds this result is
suggestive rather than firmly established.

\subsection{Spectral Verification of Suppression}
\label{sec:spectral}
\begin{figure}[H]
\centering
\includegraphics[width=0.8\textwidth]{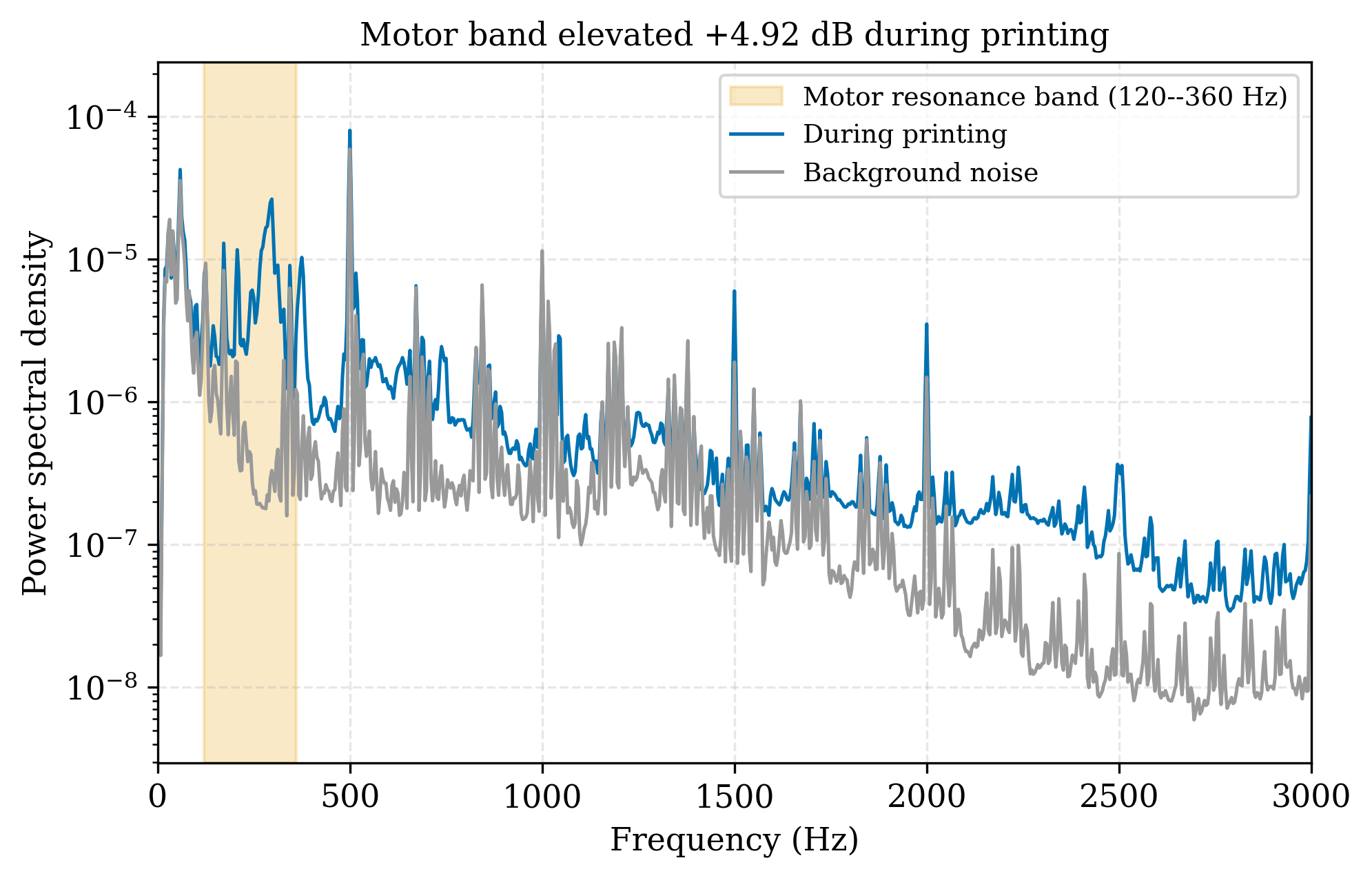}
\caption{Power spectral density during printing versus background, with the
120--360\,Hz motor-resonance band shaded.}
\label{fig:spectrum}
\end{figure}

The preceding results are classifier-based and cannot by themselves establish that the
suppression mechanism is operating. We therefore compared print audio against the
dataset's own background noise recordings in the motor-resonance band
(120--360\,Hz), referenced to a 600--2000\,Hz band. During printing the motor band
sits 9.25\,dB above the reference; in background recordings the same ratio is
4.33\,dB. The motor band is therefore elevated by only 4.92\,dB during active printing
(Figure~\ref{fig:spectrum}).

This is direct physical evidence that motor-band emission is being attenuated,
independent of classification accuracy, and it is the closest available substitute for
an on/off ablation the hardware does not permit. Suppression is operating, and
information survives it.

\FloatBarrier

\section{Discussion: Security Implications}

\textbf{Acoustic suppression buys observation time, not immunity.} A defender who
treats an acoustic noise-reduction feature as a side-channel control is protected only
against an adversary constrained to short observation windows. At 30 seconds the
channel appears dead; at 60 seconds it is well above chance, and distributed sampling
across the print raises it further. The correct security statement is a threshold, not
a binary.

\textbf{The dominant channel has no acoustic mitigation.} Print duration discriminates
better than any sensor channel here, requires no proximity, installs nothing, and
leaves no artifact to discover. Neither acoustic suppression nor vibration isolation
addresses it. Mitigation would require obscuring start and stop times---for example
randomized queuing delays or dummy print cycles---which trades directly against
throughput and scheduling transparency. We note the scope condition: duration
discriminates this strongly partly because these 12 objects span a 96-fold range in
print time. Against a catalog of similarly-sized parts its power would be lower.

\textbf{The threat is bounded.} What is demonstrated is closed-set
identification---confirming which of a set of known designs is on the bed---not
reconstruction of unknown geometry. This supports monitoring or espionage against a
known design catalog, not theft of arbitrary intellectual property.

\subsection{Real-World Applicability and Detectability}
The access an adversary requires is lower than the mounted-sensor framing implies. A
consumer handset in proximity performs comparably to an accelerometer bolted to the
machine (Section~\ref{sec:persystem}), and the dominant channel---print
duration---needs no sensor at all, only the times at which a job starts and stops.

The two configurations differ sharply in detectability. An installed accelerometer and
its microcontroller are physical artifacts, discoverable on inspection and
attributable once found. A handset resting on a bench is not distinguishable from
ordinary presence, and observation of start and stop times leaves no artifact
whatsoever. Defences that rely on detecting instrumentation therefore address only the
weaker of the two configurations.

Realistic settings are those in which unattended machines run within reach of
transient occupants: shared makerspaces, university teaching laboratories, and
co-located tenancy in shared facilities. Lower-access variants we did not test---a
device resting on the same bench rather than the machine, floor-coupled sensing, or
standoff laser vibrometry---would extend the threat further, and standoff attenuation
and through-surface coupling remain open questions.

\textbf{Device heterogeneity is a defensive lever.} Vibration signatures do not
transfer across architectures, and on the bed-slinger device vibration is not
distinguishable from chance at all. A heterogeneous fleet imposes per-device
calibration cost on an adversary that a homogeneous fleet does not.

\section{Limitations}

\textbf{Scope.} Two FDM printers from a single manufacturer, 144 recordings, 12 object
classes. Results are scoped to within-dataset conditions.

\textbf{No causal measurement of suppression.} AMNC is active throughout collection
with no user-accessible off state. We report leakage in its presence and spectral
evidence that it operates; we do not measure what it reduces.

\textbf{Observable bandwidth.} Measured vibration sampling rates of 500.0\,Hz and
199.6\,Hz correspond to Nyquist limits of 250\,Hz and 99.5\,Hz. Structural resonances
in FDM printers occur above these limits. Our finding that frequency-domain features
carry no information therefore establishes that waveform shape above the observable
bandwidth was not captured, not that shape is uninformative in principle.

\textbf{Duration matching is by observation window.} The T2 control equalizes the
observation window in seconds. It does not verify that no residual duration
information survives in the truncated features, and the direction of that error would
reduce, not inflate, vibration's apparent independent contribution.

\textbf{G-code ground truth.} The sliced G-code files distributed with the dataset are
from the P1P; those print times were mapped onto A1~Mini recordings as well, which
likely attenuates the observed correlation.

\textbf{Statistical power.} Twelve recordings per class bounds the precision of
per-stratum estimates, and the temporal result rests on five seeds.

\textbf{Task.} The task is closed-set identification; no geometric reconstruction is
demonstrated.

\section{Future Work}
\label{sec:future}
(1)~\emph{Duration-controlled replication.} The methodological finding here---that
print duration can dominate reported side-channel accuracy---should be tested against
prior results on object sets with narrow duration ranges.
(2)~\emph{Non-suppressed baseline.} Running an identical pipeline on a printer without
active acoustic suppression would quantify what suppression reduces, which this
dataset cannot.
(3)~\emph{Higher-bandwidth vibration capture.} Sampling above the structural resonance
range would test whether waveform shape carries geometry once it is observable.
(4)~\emph{Duration mitigation.} Randomized queuing, dummy cycles, and batch scheduling
should be evaluated for their cost in throughput against their leakage reduction.
(5)~\emph{Reconstruction.} Sequence-to-sequence regression from vibration onto the
toolpath, with the G-code ground truth this dataset provides, is the proper test of
whether identification generalizes to
recovery~\cite{garza2025,jamarani2025,gatlin2021}.

\section{Conclusion}
Active noise cancellation measurably attenuates motor-band acoustic emissions, and
side-channel attacks survive it. Acoustic classification is at chance only for short
observation windows, reaching 27.08\% once an adversary observes a full minute of a
print. The strongest discriminator in this dataset is not an acoustic or vibration
signature at all, but print duration, which classifies at 63.89\% from recording length
alone, tracks sliced G-code print time at \(r=0.907\), and is unaddressed by any
acoustic countermeasure. Vibration carries genuine geometry-dependent information
under equalized observation, though it adds nothing measurable beyond duration, and it
is architecture-specific to the point of vanishing on one of the two machines studied.
A consumer handset recovers as much as instrumented hardware. Securing additive
manufacturing against design disclosure, therefore, requires attention to timing and
scheduling channels, not acoustic emission alone. We release all code and analysis
artifacts.

\section*{Data and Code Availability}
Dataset: Zenodo DOI 10.5281/zenodo.13329934. Code:
\codeurl{https://github.com/ericyoc/3d_printer_sidechannel_poc}.

\bibliographystyle{unsrt}
\bibliography{references}

\end{document}